\documentclass[12pt]{article}

\usepackage{graphics}
\usepackage{epsfig}
\usepackage{cite}
\input epsf

\title{Testing for $k_T$-factorization with inclusive prompt photon production at LHC}

\author{A.V.~Lipatov, M.A.~Malyshev, N.P.~Zotov}

\begin{document}

\maketitle

\begin{center}

{\it D.V.~Skobeltsyn Institute of Nuclear Physics,\\ 
M.V. Lomonosov Moscow State University,
\\119991 Moscow, Russia\/}\\[3mm]

\end{center}

\vspace{0.5cm}

\begin{center}

{\bf Abstract }

\end{center}

We study the possibility to analyse the data on inclusive prompt photon production at first LHC runs 
in the framework of the $k_T$-factorization QCD approach. 
Our consideration is based on the amplitude for the production of a single
photon associated with a quark pair in the fusion of two off-shell gluons.
The quark component is taken into account separately using the quark-gluon Compton scatterring
and quark-antiquark annihilation QCD subprocesses.
The unintegrated parton densities in a proton are determined using
the Kimber-Martin-Ryskin (KMR) prescription as well as the CCFM evolution equation.

\vspace{0.8cm}

\noindent
PACS number(s): 12.38.-t, 13.85.-t

\newpage
The present note is motivated by the very 
recent measurements~\cite{1,2} performed by the 
CMS and ATLAS collaborations at the LHC where the isolated photon cross sections 
have been presented as functions of their transverse momentum
at $\sqrt s = 7$~TeV for the first time.

The production of prompt photons\footnote{Usually the photons are called ”prompt” if they are
coupled to the interacting quarks.} in hadron-hadron collisions at high energies
is a subject of pointed discussions up to now.
The theoretical and experimental investigations of such
processes provide a direct probe of the hard subprocess dynamics since the produced
photons are largely insensitive to the effects of final-state hadronization. 
The cross sections of these processes are strongly sensitive to the
parton (quark and gluon) content of a proton since
isolated prompt photons can be produced mainly 
via quark-gluon Compton scattering or quark-antiquark
annihilation (at LO).

We note, however, that a complete theoretical description of the Tevatron
data within the QCD is an open question still (see~\cite{3, 4}
and references therein).
It was demonstrated~\cite{5, 6} that the overall description of these data in the 
framework of the standard QCD can be 
achieved by introducing some additional intrinsic transverse momentum $k_T$ of 
incoming partons, which is usually assumed to have a Gaussian-like distribution. 
The average value of this $k_T$ increases from $k_T \sim 1$~GeV to more than 
$k_T \sim 3$~GeV in hard scattering processes as the $\sqrt s$ increases from UA6 to 
Tevatron energies~\cite{5, 7}. The importance of including the gluon emission through the 
resummation formalism was recognized and only recently this approach has been developed 
for inclusive prompt photon production~\cite{8,9,10,11}.

In the framework of the $k_T$-factorization approach of QCD~\cite{12}, 
which is of primary consideration in this note, the transverse momentum of incoming partons 
occurs in a natural way. In this approach, the non-zero partonic 
$k_T$ is generated perturbatively in the course of non-collinear parton evolution via the 
corresponding (usually BFKL~\cite{13} or CCFM~\cite{14}) evolution equations.
A detailed description of the $k_T$-factorization can be
found, for example, in reviews~\cite{15}. Studies of the prompt photon
production at hadronic colliders in the framework of $k_T$-factorization approach 
have been done in many papers~\cite{16,17,18,19,20}. So, investigations~\cite{16,17,18} have been
based on the leading-order matrix elements of quark-gluon Compton scattering 
and quark-antiquark annihilation subprocesses.
An important component of these calculations is
the unintegrated ($k_T$-dependent) quark distributions in a proton. 
At present, these densities are available in
the framework of Kimber-Martin-Ryskin (KMR)~\cite{21} approach only since
there are some theoretical difficulties
in obtaining the quark distributions immediately from CCFM or BFKL equations 
(see, for example,~\cite{15} for more details).
In our previous investigations~\cite{19,20} we tried an another way.
The main idea of~\cite{19,20} was in reexpressing the quark 
contributions in terms of 
gluon ones using the higher-order off-shell 
gluon-gluon fusion matrix elements, namely $g^* + g^* \to \gamma + q + \bar q$.
Thus, we reduced the problem of poorly known and poorly 
calculable unintegrated quark distributions to much better investigated gluon densities.
The corresponding contributions from the valence quarks have been taken into account 
separately. We obtained~\cite{19,20} a reasonably good agreement between the $k_T$-factorization 
predictions and the Tevatron data on the inclusive prompt photon production
cross cross sections in both central and forward photon pseudo-rapidity regions. 
Based on the results~\cite{19,20}, in the present note we give the first systematic analysis of data~\cite{1,2}
taken by the CMS and ATLAS collaborations at the LHC in the framework of the $k_T$-factorization approach. 
Moreover, we improve our previous predictions~\cite{19,20} by taking into
account the transverse momentum of incoming off-shell quarks in a proper way.
In other aspects we will strictly follow the approach
described in~\cite{19,20}.

  We only briefly recall here
the corner-stones of proposed theoretical scheme.
The starting point of consideration is the usual leading order 
${\cal O}(\alpha)$ and ${\cal O}(\alpha \alpha_s)$ off-shell subprocesses, namely
$q^* + g^* \to \gamma + q$ and $q^* + \bar q^* \to \gamma + g$.
These subprocesses are strongly depend on the unintegrated quark densities in a proton,
$f_q(x,{\mathbf k}_{T}^2,\mu^2)$. The unintegrated quark distributions include the ones of valence quarks $f_q^{(v)}(x,{\mathbf k}_{T}^2,\mu^2)$, sea quarks
appearing at the last step of the gluon evolution $f_q^{(g)}(x,{\mathbf k}_{T}^2,\mu^2)$
and sea quarks coming from the earlier (second-to-last, third-to-last and other) 
gluon splittings $f_q^{(s)}(x,{\mathbf k}_{T}^2,\mu^2)$.
In the proposed approach~\cite{19,20} we simulate the last gluon splitting by the higher-order 
${\cal O}(\alpha \alpha_s^2)$ off-shell matrix elements 
$g^* + g^* \to \gamma + q + \bar q$. In this way we take into account the 
contributions from the $f_q^{(g)}(x,{\mathbf k}_T^2,\mu^2)$.
To estimate the contributions from 
the $f_q^{(v)}(x,{\mathbf k}_{T}^2,\mu^2)$ and $f_q^{(s)}(x,{\mathbf k}_{T}^2,\mu^2)$
we use the specific properties of the KMR formalism which 
enables us to discriminate between the various components of the unintegrated quark densities (see below).

Thus, the proposed scheme results to the following partonic 
subprocesses\footnote{We will neglect the contributions from the so-called
fragmentation mechanisms. It is because after applying the isolation cut (see~\cite{1,2}) 
these contributions amount only to about 10\% of the visible cross section. 
The isolation requirement and additional conditions
which preserve our calculations from divergences have been specially discussed in~\cite{19,20}.}:
$$
  g^* + g^* \to \gamma + q + \bar q, \eqno(1)
$$
$$
  q^{*(v,s)} + g^* \to \gamma + q, \eqno(2)
$$
$$
  q^{*(v,s)} + \bar q^{*(v,s)} \to \gamma + g. \eqno(3)
$$

\noindent
To be precise, the gluon-gluon fusion~(1) takes into account the contribution of the  $q^{*(g)} + \bar q^{*(g)}$ annihilation subprocesse,
and the valence and sea quark-gluon scattering (2) take into
account the 
$q^{*(v)} + \bar q^{*(g)}$ and $q^{*(s)} + \bar q^{*(g)}$ mechanisms.
Evaluation of the off-shell matrix elements of subprocesses (1) --- (3)
is straightforward and the analytical expressions have been listed in~\cite{19}. 
Here we only would like to mention two technical points.
First, according to the $k_T$-factorization prescription~\cite{12},
the summation over the incoming off-shell gluon polarizations in (1) and (2) is 
carried with $\sum \epsilon^\mu \epsilon^{  \, \nu} = {\mathbf k}_T^{\mu} {\mathbf k}_T^{\nu}/{\mathbf k}_T^2$,
where ${\mathbf k}_T$ is the gluon transverse momentum.
Second, when we calculate the matrix element squared,
the spin density matrix for off-shell spinors in (2) and (3) is taken in the
form $u (p) \bar u (p) = x \hat p_p$~\cite{22}, where $x$ is the fraction of 
initial proton longitudinal momentum $p_p$. In all other respects our calculations follow the standard Feynman rules.
Since the expression for the off-shell quark spin density matrix has been derived
in the massless approximation, numerically we neglect the charmed quark mass. 

According to the $k_T$-factorization theorem, to calculate the cross section of the prompt photon production
one should convolute the off-shell partonic cross sections (1) --- (3)
with the relevant unintegrated quark and/or gluon distributions in a proton:
$$
  \sigma = \sum_{i,j = q,\,g} \int {\hat \sigma}_{ij}^*(x_1, x_2, {\mathbf k}_{1T}^2, {\mathbf k}_{2T}^2) \, f_i(x_1,{\mathbf k}_{1T}^2,\mu^2) f_j(x_2,{\mathbf k}_{2T}^2,\mu^2) \, dx_1 dx_2 \, d{\mathbf k}_{1T}^2 d{\mathbf k}_{2T}^2, \eqno(4)
$$

\noindent
where ${\hat \sigma}_{ij}^*(x_1, x_2, {\mathbf k}_{1T}^2, {\mathbf k}_{2T}^2)$
is the relevant partonic cross section. The initial off-shell partons have fractions $x_1$ and $x_2$ of 
initial protons longitudinal momenta and non-zero transverse momenta 
${\mathbf k}_{1T}$ and ${\mathbf k}_{2T}$.
The analytical expressions for the contributions of subprocesses (1) --- (3) are given in~\cite{19}.

Concerning the unintegrated parton densities, we use
the CCFM-evolved gluon and valence quark distributions
derived in~\cite{23} and~\cite{24}, respectively.
To determine unintegrated parton densities we also apply
the KMR approximation~\cite{21}.
The KMR approach is the formalism to construct the unintegrated parton distributions
$f_a(x,{\mathbf k}_T^2,\mu^2)$ from the known conventional parton
distributions $xa(x,\mu^2)$, where $a = g$ or $a = q$. 
In this approximation, the unintegrated quark and 
gluon distributions are given by~\cite{21}
$$
  \displaystyle f_q(x,{\mathbf k}_T^2,\mu^2) = T_q({\mathbf k}_T^2,\mu^2) {\alpha_s({\mathbf k}_T^2)\over 2\pi} \times \atop {
  \displaystyle \times \int\limits_x^1 dz \left[P_{qq}(z) {x\over z} q\left({x\over z},{\mathbf k}_T^2\right) \Theta\left(\Delta - z\right) + P_{qg}(z) {x\over z} g\left({x\over z},{\mathbf k}_T^2\right) \right],} \eqno (5)
$$
$$
  \displaystyle f_g(x,{\mathbf k}_T^2,\mu^2) = T_g({\mathbf k}_T^2,\mu^2) {\alpha_s({\mathbf k}_T^2)\over 2\pi} \times \atop {
  \displaystyle \times \int\limits_x^1 dz \left[\sum_q P_{gq}(z) {x\over z} q\left({x\over z},{\mathbf k}_T^2\right) + P_{gg}(z) {x\over z} g\left({x\over z},{\mathbf k}_T^2\right)\Theta\left(\Delta - z\right) \right],} \eqno (6)
$$

\noindent
where $P_{ab}(z)$ are the usual unregulated LO DGLAP splitting 
functions. The theta functions which appear 
in~(5) and~(6) imply the angular-ordering constraint $\Delta = \mu/(\mu + |{\mathbf k}_T|)$ 
specifically to the last evolution step to regulate the soft gluon
singularities. For other evolution steps, the strong ordering in 
transverse momentum within the DGLAP equations automatically 
ensures angular ordering\footnote{Numerically, in (5) and (6) we have applied the 
Gl\"uck-Reya-Vogt (GRV) parton distributions~\cite{25}.}.
The Sudakov form factors $T_q({\mathbf k}_T^2,\mu^2)$ and 
$T_g({\mathbf k}_T^2,\mu^2)$ which appears in (5) and (6) enable us to include logarithmic loop corrections
to the calculated cross sections. 

Note that the function $f_q(x,{\mathbf k}_T^2,\mu^2)$ in~(5)
represents the total quark distribution function in a proton.
Modifying~(5) in such a way that only the first term is 
kept and the second term omitted, we switch the last gluon splitting 
off, thus excluding the $f_q^{(g)}(x,{\mathbf k}_T^2,\mu^2)$ component.
Taking the difference between the quark and antiquark densities we
extract the valence quark component $f_q^{(v)}(x,{\mathbf k}_T^2,\mu^2)=
f_q(x,{\mathbf k}_T^2,\mu^2)-f_{\bar{q}}(x,{\mathbf k}_T^2,\mu^2)$.
Finally, keeping only sea quark in first term of~(5) we remove the 
valence quarks from the evolution ladder.
In this way only the $f_q^{(s)}(x,{\mathbf k}_T^2,\mu^2)$ contributions to the 
$f_q(x,{\mathbf k}_T^2,\mu^2)$ 
are taken into account. 

Other essential parameters were taken as follows:
renormalization and factorization scales $\mu = \xi E_T^\gamma$ (where
we vary the parameter $\xi$ between 1/2 and 2 about the default value $\xi = 1$
in order to estimate the scale uncertainties of our calculations),
LO formula for the strong coupling constant $\alpha_s(\mu^2)$ 
with $n_f = 4$ massless quark flavours and $\Lambda_{\rm QCD} = 200$ MeV, 
such that $\alpha_s(M_Z^2) = 0.1232$.
The multidimensional integration in~(4) has been performed
by means of the Monte Carlo technique, using the routine \textsc{vegas}~\cite{26}.
The full C$++$ code is available from the authors on 
request\footnote{lipatov@theory.sinp.msu.ru}. This code is practically 
identical to that used in~\cite{19,20} with exception that now we apply it to the LHC conditions.

The results of our calculations are shown in Fig.~1 where we confront
the calculated differential cross section of the inclusive isolated prompt photon production
(as a function of the photon transverse energy $E_T^\gamma$) with the first LHC data~\cite{1,2}
taken by the CMS and ATLAS collaborations at $\sqrt s = 7$~TeV.
In Fig.~2 we show the data/theory ratio of our predictions.
The CMS data refer to the kinematic region defined by $E_T^\gamma > 21$~GeV and $|\eta^\gamma| < 1.45$.
The ATLAS data have been measured at $15 < E_T^\gamma < 100$~GeV,  $|y^\gamma| < 0.6$,
$0.6 < |y^\gamma| < 1.37$ and $1.52 < |y^\gamma| < 1.81$, respectively.
In Figs.~1 and 2, the solid and dash-dotted histograms are obtained with the KMR and CCFM parton densities by fixing both the
factorization and normalization scales at the default value $\mu = E_T^\gamma$. 
The upper and lower dashed histograms correspond to the scale variation in KMR predictions as it
was described above. 
One can see that the predictions obtained with the KMR partons tend to slightly overestimate the measured cross sections, but
they are coincide with the data within the theoretical uncertainties.
Contrary, predictions based on the CCFM evolution lie below the LHC data in a wide $E_T^\gamma$ range.
The main reason of such behaviour is connected with taking into account the contributions from the sea quarks
originating from the earlier steps of the evolution cascade
in the former case. However, to avoid the possible double counting
these contributions are not taken into account in the
CCFM calculations since part of them can be already included into the CCFM results (via initial 
parton distributions which enter to the CCFM equation).
Note that the similar effect has been observed~\cite{22} in the prompt photon photoproduction at HERA.

The relative contributions of subprocesses (1) --- (3) to the prompt photon cross section 
are shown in Fig.~3. Here we use the KMR parton densities for illustration. 
One can see that at low $E_T$ the gluon fusion subprocess gives the dominant contribution
whereas the QCD Compton and quark-antiquark annihilation subprocesses are mostly significant at higher $E_T$. 
The contribution from the quarks appearing from the earlier evolution steps, $f_q^{(s)}(x,{\mathbf k}_{T}^2,\mu^2)$,
is also significant in both central and more forward (as it was measured by the ATLAS collaboration) rapidities.


To conclude, in the present note we apply the CCFM and KMR unintegrated parton densities
to the analysis of the first experimental data on the prompt photon
production in $pp$ collisions taken by the CMS and ATLAS collaborations at the LHC.
We have obtained a reasonably good agreement between our predictions and the LHC data and have  
demonstrated that the quarks are important even at the LHC energies and therefore 
should be properly included into the non-collinear evolution equations. 
It is important for further studies of small-$x$ physics at the LHC, in particular, 
for investigation of Drell-Yan pair production which give us possibilities to test
the region of very low $x$ (up to $x \sim 10^{-5}$).

{\sl Acknowledgements.} 
We thank convener of the ATLAS Standard Model Direct Photon working group
Marco Delmastro for his interest and very useful remarks.
A.V.L and N.P.Z. are very grateful to 
DESY Directorate for the support in the 
framework of Moscow --- DESY project on Monte-Carlo
implementation for HERA --- LHC. 
A.V.L. and M.A.M. were supported in part by the grant of the president of 
Russian Federation (MK-3977.2011.2).
Also this research was supported by the 
FASI of Russian Federation (grant NS-4142.2010.2) and 
FASI state contract 02.740.11.0244.

\newpage


\begin{figure}
\begin{center}
\epsfig{figure=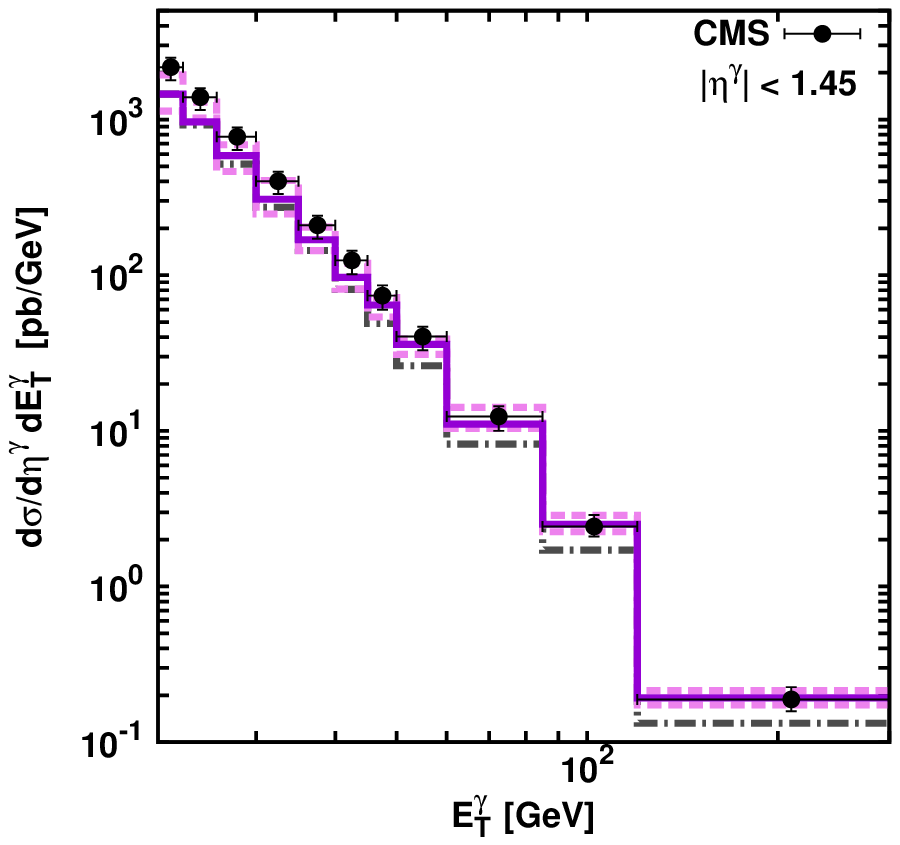, width = 8.1cm}
\epsfig{figure=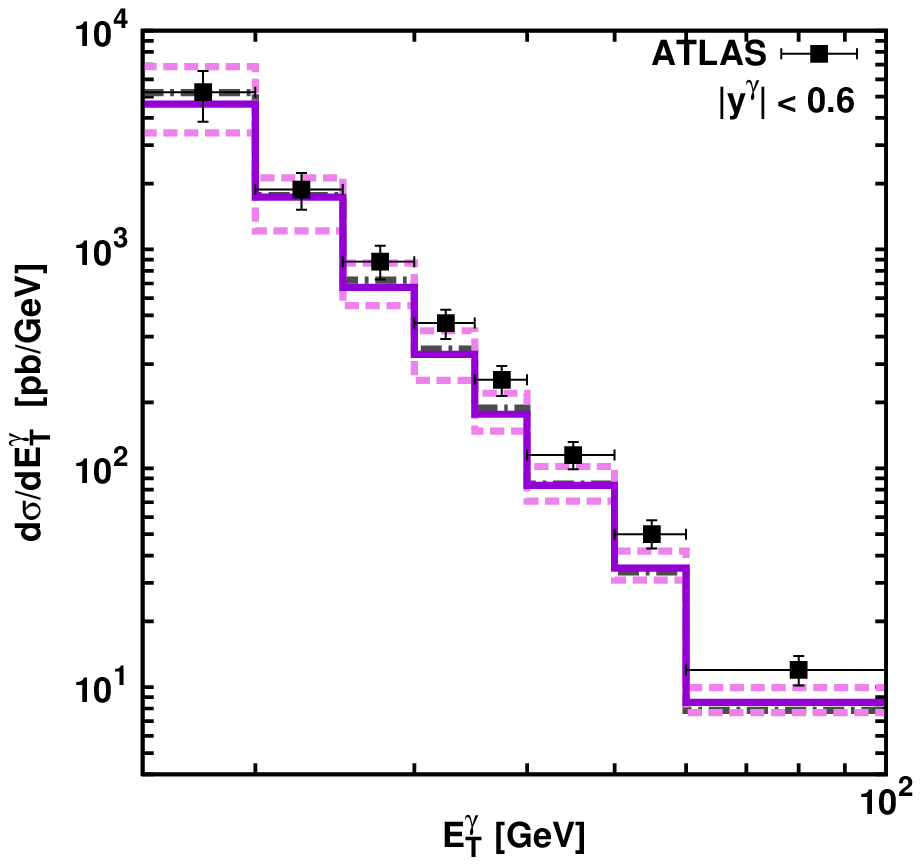, width = 8.1cm}
\epsfig{figure=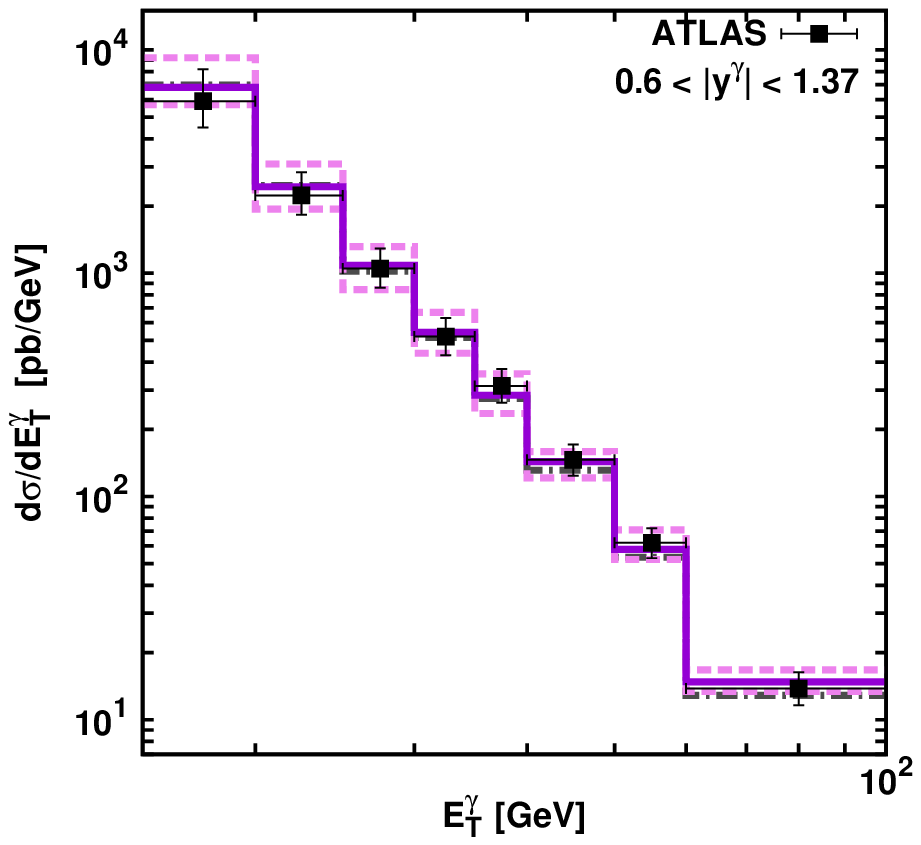, width = 8.1cm}
\epsfig{figure=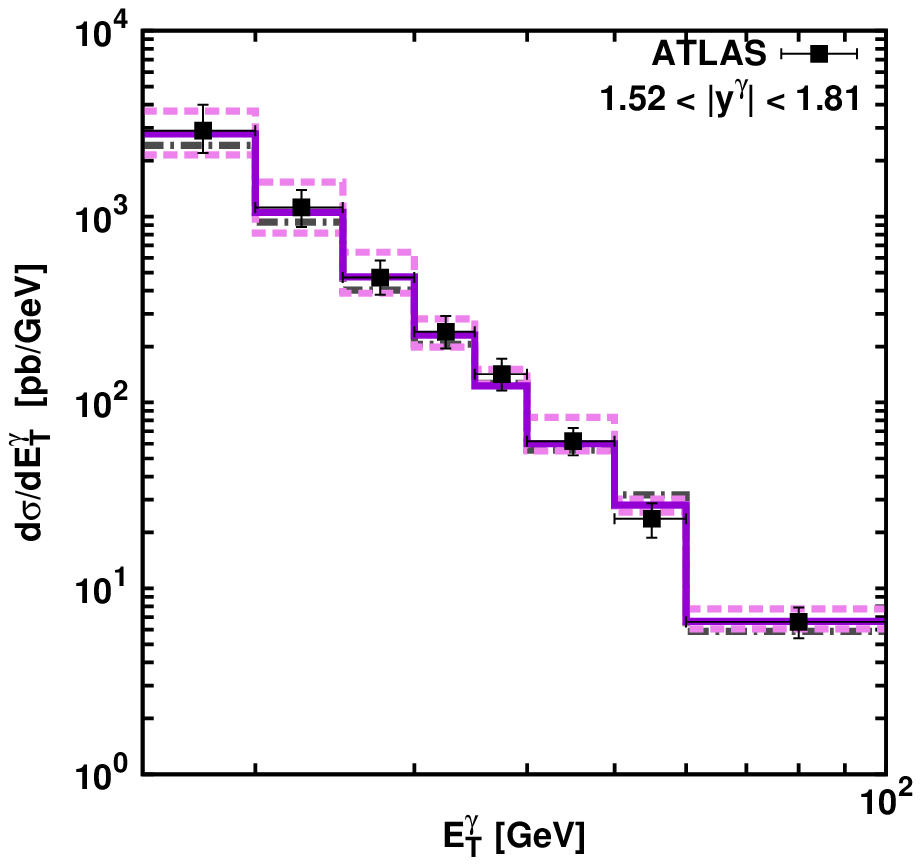, width = 8.1cm}
\caption{The differential cross sections of the 
inclusive prompt photon production in $pp$ collisions 
as a function of $E_T^\gamma$ calculated
at $\sqrt s = 7$~TeV.
The solid histogram corresponds to the KMR predictions at the 
default scale $\mu = E_T^\gamma$, whereas the upper and 
lower dashed histograms correspond to scale variations described in the text.
The dash-dotted histogram corresponds to the predictions obtained with the CCFM 
parton densities. The experimental data are from CMS~\cite{1} and ATLAS~\cite{2}.}
\end{center}
\label{fig1}
\end{figure}

\newpage

\begin{figure}
\begin{center}
\epsfig{figure=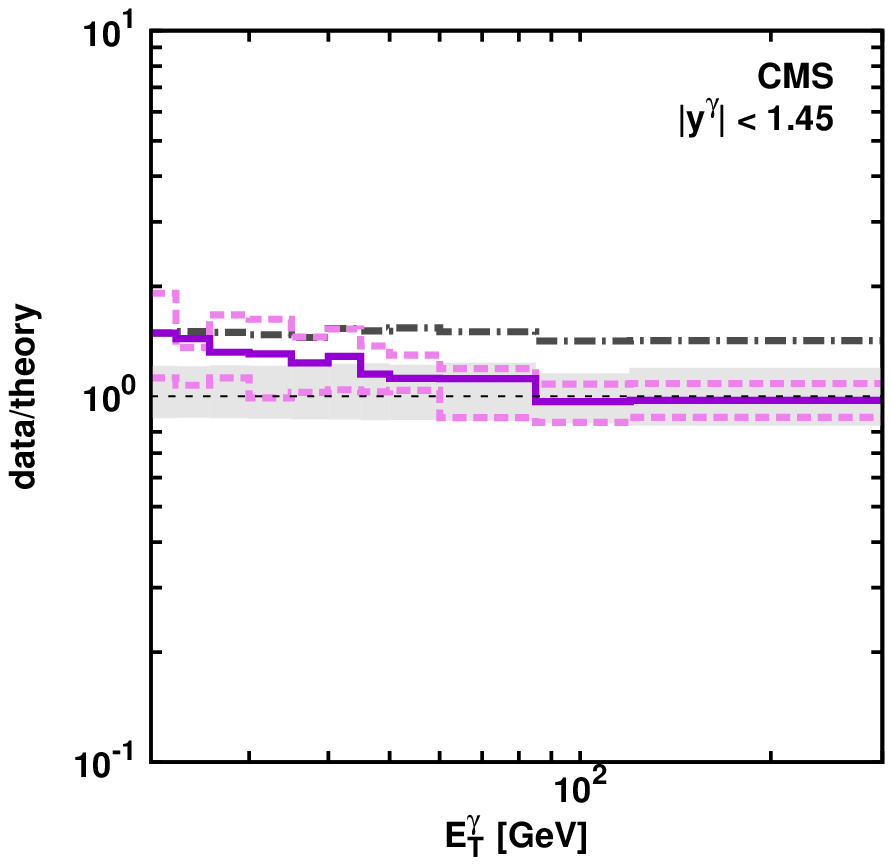, width = 8.1cm}
\epsfig{figure=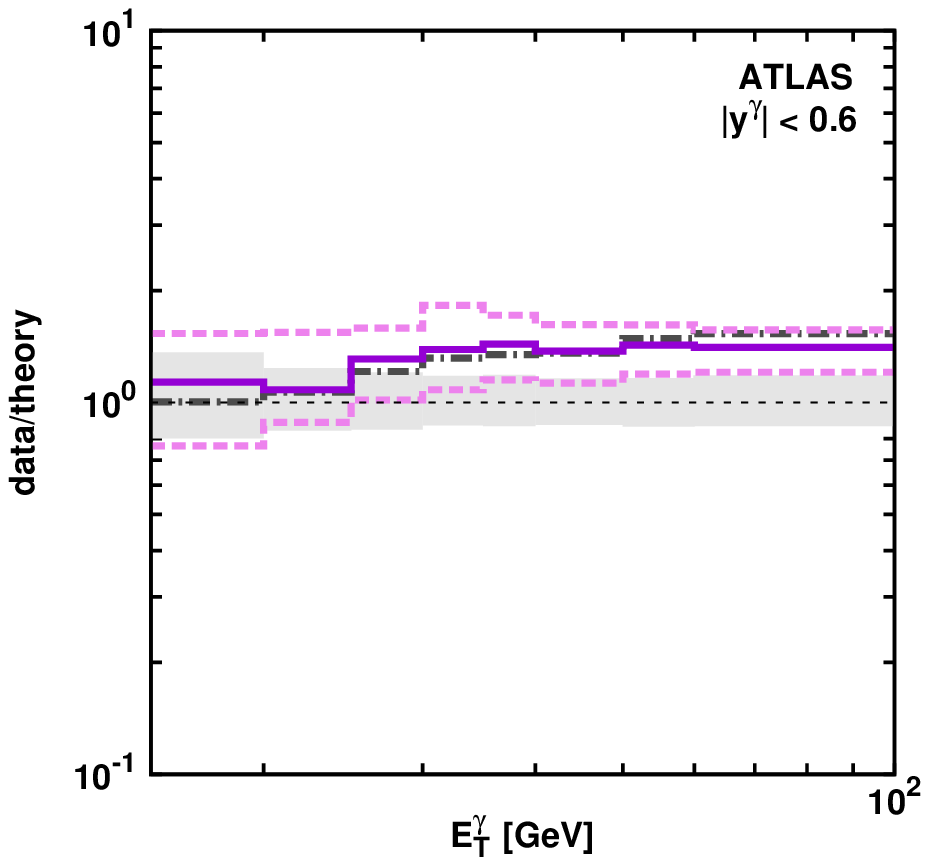, width = 8.1cm}
\epsfig{figure=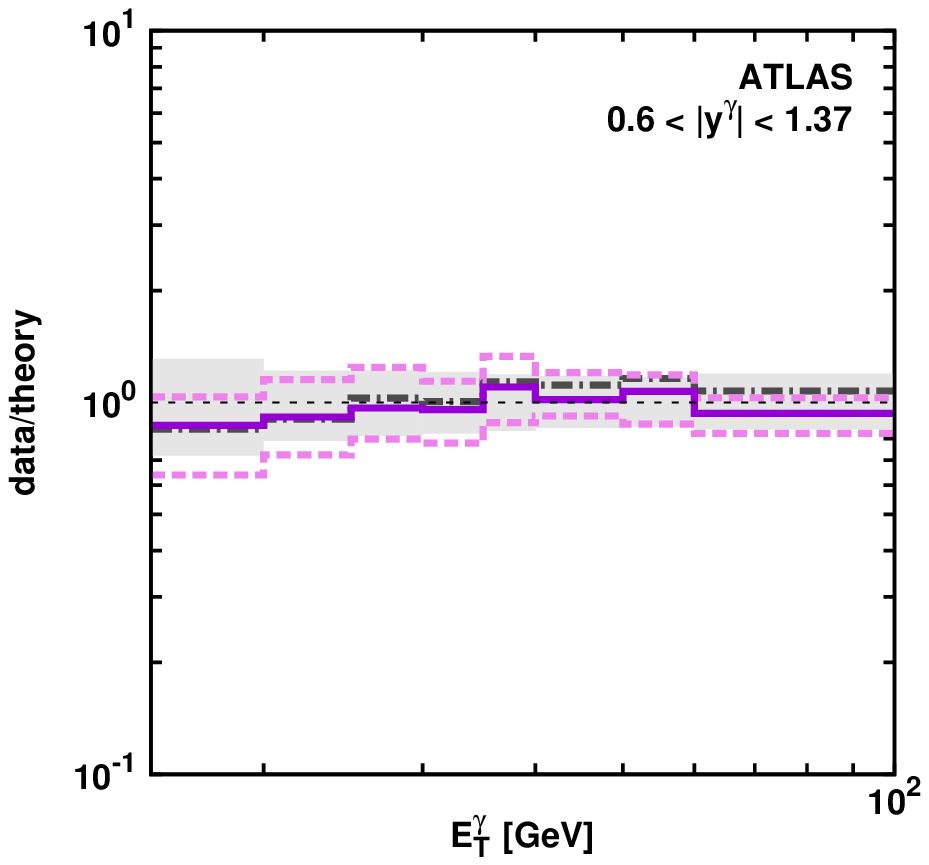, width = 8.1cm}
\epsfig{figure=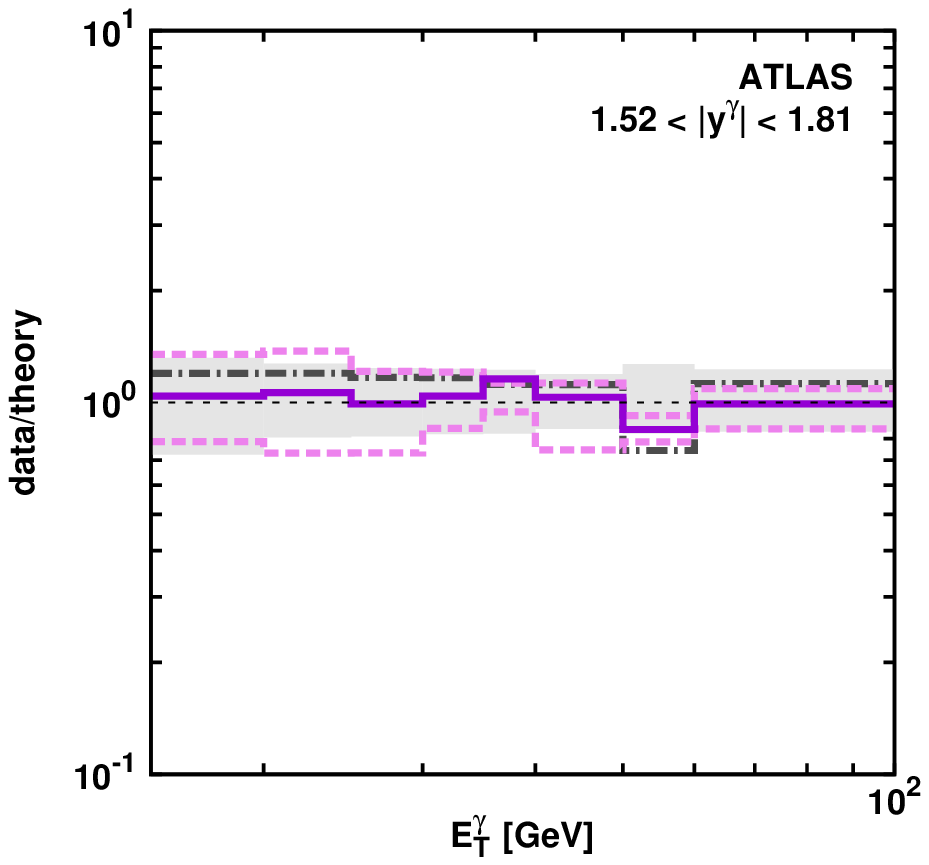, width = 8.1cm}
\caption{The theory/data ratio of the 
inclusive prompt photon production in $pp$ collisions 
as a function of $E_T^\gamma$ calculated
at $\sqrt s = 7$~TeV. Notation of the histograms is 
the same as in Fig.~1. The shaded band represent the 
experimental uncertainties.}
\end{center}
\label{fig2}
\end{figure}

\newpage

\begin{figure}
\begin{center}
\epsfig{figure=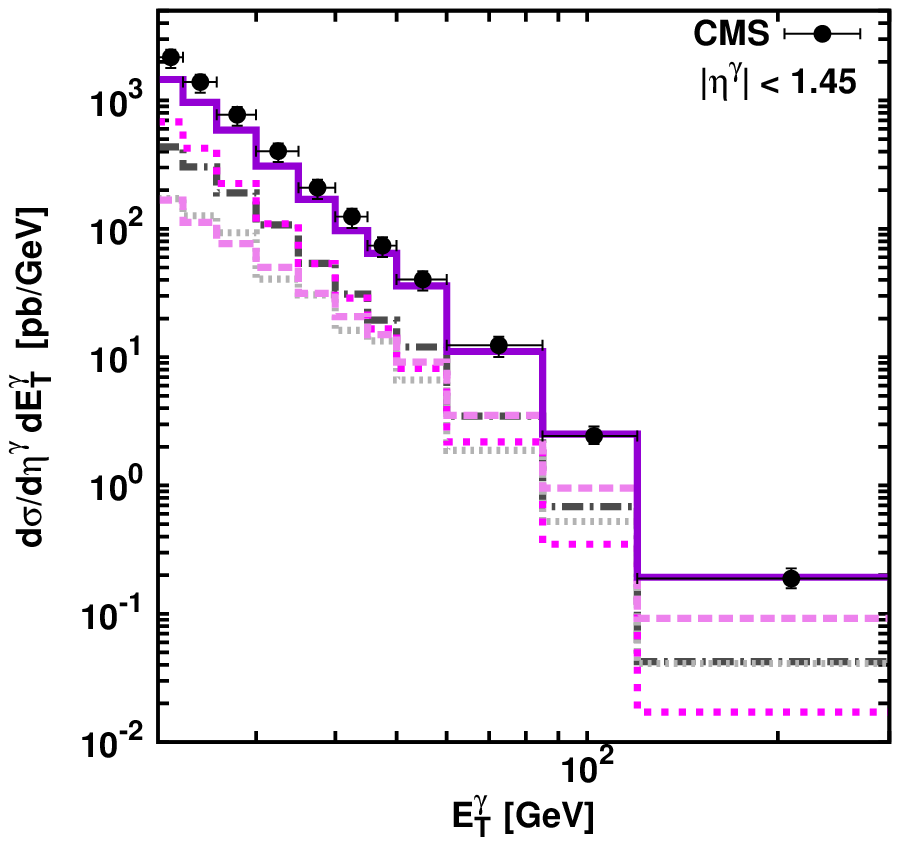, width = 8.1cm}
\epsfig{figure=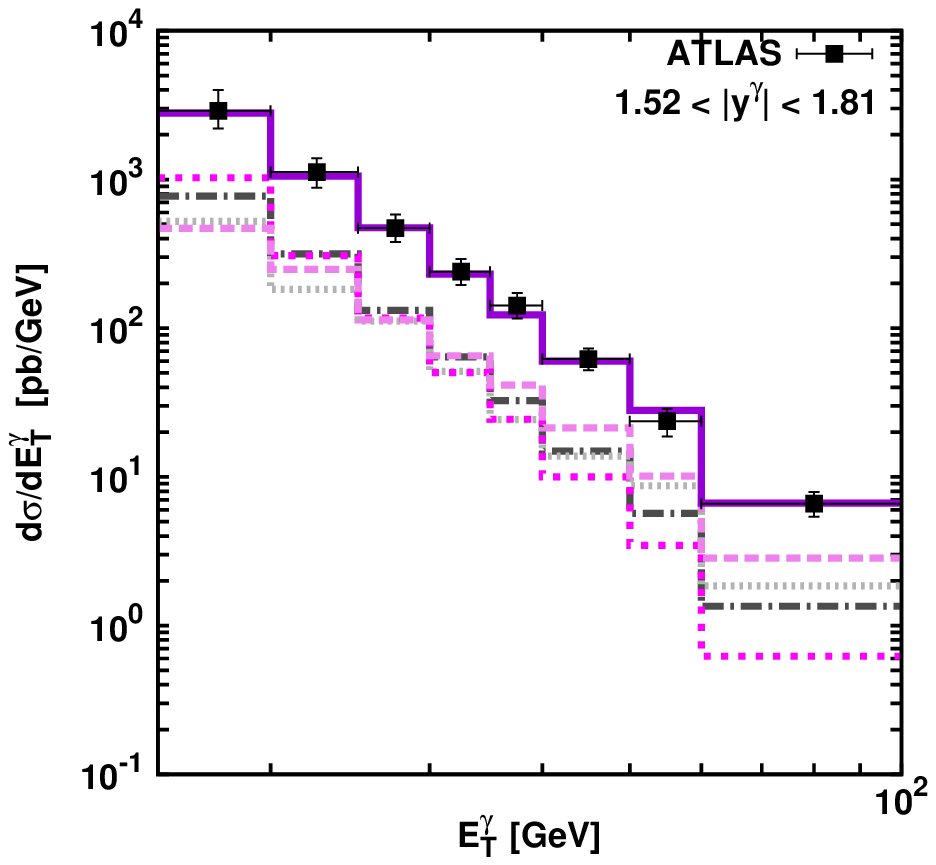, width = 8.1cm}
\caption{Different contributions to the cross section of the 
inclusive prompt photon production in $pp$ collisions 
as a function of $E_T^\gamma$ calculated
at $\sqrt s = 7$~TeV.
The dashed, short dashed and dotted histograms correspond to the
contributions from the $q^*g^*\to q\gamma$, $g^*g^*\to q\bar q\gamma$ and $q^*\bar q^*\to g\gamma$
subprocesses, respectively. The dash dotted histogram corresponds to the 
$f_q^{(s)}(x,{\mathbf k}_{T}^2,\mu^2)$ contribution.
The solid histogram represents the sum of all these components.
We use the KMR parton densities for illustration. 
The experimental data are from CMS~\cite{1} and ATLAS~\cite{2}.}
\label{fig3}
\end{center}
\end{figure}

\end{document}